\documentclass[aps,shownopacs,superscriptaddress,nofootinbib,preprint,11pt]{revtex4}
\usepackage{hyperref}
\usepackage{amsmath}
\usepackage{float}
\usepackage{graphics}
\usepackage{graphicx}
\usepackage{epsfig}
\usepackage{psfrag}
\usepackage{color}
\usepackage{slashed}
\usepackage{multirow}

\usepackage{amsfonts}
\usepackage{amssymb}

\newcommand*{\be}{\begin{equation}}
	\newcommand*{\ee}{\end{equation}}
\newcommand*{\bea}{\begin{eqnarray}}
	\newcommand*{\eea}{\end{eqnarray}}

\newcommand{\comment}[1]{}

\newcommand{\cref}[1]{Chapter~\ref{c.#1}}

\def\beq{\begin{equation}}
	\def\eeq{\end{equation}}
\def\bea{\begin{eqnarray}}
	\def\eea{\end{eqnarray}}
\def\ba{\begin{array}}
	\def\ea{\end{array}}
\def\bi{\begin{itemize}}
	\def\ei{\end{itemize}}
\def\be{\begin{enumerate}}
	\def\ee{\end{enumerate}}
\def\bc{\begin{center}}
	\def\ec{\end{center}}
\def\bt{\begin{table}}
	\def\et{\end{table}}
\def\btb{\begin{tabular}}
	\def\etb{\end{tabular}}

	\def\lsim{\raise0.3ex\hbox{$\;<$\kern-0.75em\raise-1.1ex\hbox{$\sim\;$}}}
	\def\gsim{\raise0.3ex\hbox{$\;>$\kern-0.75em\raise-1.1ex\hbox{$\sim\;$}}}

\begin{document}

		\title{Bulk RS models, Electroweak Precision tests and the 125 GeV Higgs}
		\author{Abhishek M Iyer}
		\email{abhishek@theory.tifr.res.in}
		\affiliation{Department of Theoretical Physics, Tata Institute of Fundamental Research, Homi Bhabha Road, Colaba, Mumbai 400 005, India}
		\author{K. Sridhar}
		\email{sridhar@theory.tifr.res.in}
		\affiliation{Department of Theoretical Physics, Tata Institute of Fundamental Research, Homi Bhabha Road, Colaba, Mumbai 400 005, India}
		\author{Sudhir K. Vempati}
		\email{vempati@cts.iisc.ernet.in}
		\affiliation{Centre for High Energy Physics, Indian Institute of Science,
			Bangalore 560012, India}
		\begin{abstract}
			%We investigate the status of models with a Randall-Sundrum (RS)-like 
			%geometry by performing a global fit of the up-to-date measurements of
			%precision observables to the model parameters. In addition to the Bulk 
			%RS model (including Bulk Higgs), we also consider the following two model 
			%implementations: a) Models with a bulk custodial symmetry, and 
			%b) Models with a deformed metric with a deviation of the geometry 
			%from the AdS near the IR brane.
			%We find that the best fits to the data are obtained when the Higgs is not 
			%localized exactly on the IR brane. This is also consistent with the least 
			%fine-tuned scenario. The analysis of the case with a deformed metric has 
			%interesting implications for searches of Kaluza-Klien (KK) states.
			We present upto date electroweak fits of various Randall Sundrum (RS) models. 
			We consider the bulk RS model, deformed RS and the custodial RS models. 
			For the bulk RS case we find the lightest Kaluza Klein (KK) mode of the gauge boson to be $\sim 8$ TeV while for the custodial case it is $\sim 3$ TeV. 
			The deformed model is the least fine tuned of all which can give a good fit for KK masses $< 2$ TeV depending on the choice of the model parameters. 
			We also comment on the fine tuning in each case.
			
		\end{abstract}
		\vskip .5 true cm
		\preprint {TIFR/TH/15-05}
		\maketitle
		\section{Introduction}
		The discovery of the Higgs boson at $\sim 125$ GeV has firmly established the status of the Higgs mechanism as the theory of electroweak symmetry breaking physics. In addition it also fixes one of main unknown inputs of electroweak precision fits. Electroweak precision measurements put very important and sometimes very strong constraints on new physics models.
		In the present work we focus on the Randall-Sundrum 
		(RS) model \cite{RS} and its variations.
		We update the constraints on the lightest Kaluza-Klein (KK) modes of RS scenarios in the light of discovery of Higgs mass and improved measurements the W-boson mass ($m_w$) and top mass ($m_t$). Electroweak precision constraints played an important role in the evolution of RS models and their phenomenology \footnote{For a recent review see \cite{Gher2,gherghetta}}. 
		In the original standard proposal all the standard model fields are localized on the IR brane
		
		% The RS model, in its original form, 
		%was based on a slice of AdS${}_5$ spacetime between two 3-branes, referred
		%to as the UV and IR branes with a tiny inter-brane separation just 
		%somewhat larger than the Planck length. The RS geometry is warped and
		%this helps provide a natural answer to the gauge-hierarchy
		%problem by "warping down" the fundamental five-dimensional scale which
		%is of the order of $M_P$ down to the weak scale.
		%
		%In the original model, with a view to seeking a solution to the 
		%hierarchy problem, only gravity was allowed to propagate in the bulk
		%while the Standard Model (SM) fields were on the brane.
		
		Later motivated by gauge coupling unification, gauge fields were moved to the bulk while keeping the Higgs and the fermion fields on the brane \cite{Agashe:2002pr}. 
		In both cases large contributions to the oblique $S$ and $T$  parameters were noted \cite{Peskin:1991sw}.
		resulting in bounds on the first KK mass in excess of 30 TeV \cite{Davou,Hisano,Huber:2000fh,Csaki:2002gy,Burdman:2002gr,Delgado:2007ne}. 
		Moving the fermions into the bulk served the following two purposes:\newline
		\textbf{1)}~It offered an elegant solution to the Yukawa hierarchy puzzle achieved
		by localizing the fermions at different points in the bulk, resulting in 
		interesting flavour phenomenology
		in the hadronic sector \cite{AgasheSoni, Huber1,cedric,neubert1,neubert2} and the leptonic sector \cite{gross,Kitano,Huber4,Huber3,Huber2,Agashe,Fitzpatrick,Chen,AgasheSundrum,Huber1,Archer:2012qa,Iyer:2012db,Iyer:2013hca,Iyer:2013eka}. 
		For a detailed description of RS phenomenology with bulk fields see \cite{Gherghetta,Gher2}.\newline
		\textbf{2)}~The constraints on the gauge KK states from the S parameter is significantly weakened as all the light fermions except the top are localized away from the IR brane and Higgs.\newline
		The constraints from T parameter, however remain strong as the Higgs doublet is  localized near the IR brane which is necessary for the solution to the hierarchy problem.
		In view of this the following extensions were proposed:
		%\begin{itemize}
		\textbf{a} Models with bulk custodial symmetry \cite{Agashe:2003zs}
		The bulk gauge group in question $SU(2)_L\times SU(2)_R\times U(1)_X$. In this case the additional corrections to the T parameter due to new KK gauge bosons cancel the volume enhanced
		contributions due to the KK states of the SM gauge bosons.
		The T parameter vanishes at tree level and the limits on the KK mass of the first gauge boson
		is mainly due to the S parameter. A straight forward estimation of the S parameter results in a lower bound on the first KK mass to be $\sim 4$ TeV for the point to lie inside the 3$\sigma$ region in Figure[\ref{STplane}]. Taking into account the loop corrections to the T parameter, (in scenarios  with $Zbb$ protection ) it was found that one can lower the mass of the first KK gauge boson to around 3 TeV at 3 $\sigma$.
		An additional alternative to consider custodial models with gauge-Higgs unification which can address the little hierarchy problem. \cite{Carena:2006bn}. A global fit to the precision observables in such models was performed in \cite{Carena:2007ua}. Scenarios with bulk Higgs was considered in \cite{Archer:2014jca}.
		\textbf{b} Models with a deformed metric:\cite{Falkowski:2008fz,Cabrer1,Cabrer2}
		In this setup the bulk geometry is RS like (AdS) near the UV brane while there is a deviation from AdS geometry near the IR brane. Depending on the model parameters this often results
		in a smaller volume factor as compared to the original RS setup. In \cite{Cabrer3} the authors performed a fit to the data for different values of the Higgs mass and evaluated the fine tuning required to fit that particular Higgs mass. In \cite{Carmona:2011ib} the authors studied the implications of the one loop corrections to the T parameter on the fits.
		\textbf{c} Models with brane localized kinetic terms for the gauge bosons \cite{Carena:2003fx}.
		%	\end{itemize}\\
	There were several previous analyses where impact of bulk fields on oblique observables was studied. 
		With bulk fermions, the dominant constraint on the KK mass is due to the $T$ parameter which is enhanced due to the mixing of the zero mode gauge bosons with the KK modes. This mixing is governed by the Higgs vev.  As noted in \cite{Archer:2012qa},these constraints can be ameliorated with a Higgs vev not strongly localized near the IR brane which in turn reduces the zero-KK mode mixing. This set-up is particularly useful in a model with a deformed-RS metric, where the KK constraints on the lowest KK masses can be reduced to $\sim 2.5$ TeV.
		In \cite{Agashe:2013kxa} the authors while updating the bounds on the KK masses from precision electroweak data, also discuss  the impact of the future measurements of rare K and B decays on the parameter space of the model. They also discuss the correlation between the these flavour measurements and the limits from direct searches for current and future runs of the LHC. Discussions of general composite Higgs models was done in \cite{Fichet:2013ola} of which RS was considered as an example. While discussing the trilinear and quartic anomalous gauge couplings in these scenarios, they quote limits on the mass of KK gauge resonance due to precise values of the $S$ and $T$ parameters. In the absence of brane kinetic terms, the constraints on custodial models with a brane Higgs was about 7 TeV while this could be lowered to 6.6 TeV for pseudo Nambu goldstone Higgs at 95 \% CL. Models with bulk Higgs were also considered in \cite{Archer:2014jca} where a detailed analysis correlating signal strengths for different production mechanisms and decay channels was performed as function of anarchic bulk Yukawa parameter, KK masses and extent of the compositeness of the Higgs operator. In \cite{Dillon:2014zea} the authors showed that the inclusion of higher dimensional operators in the bulk and on the branes can significantly reduce the constraints on the
		T-parameter.  These operators only require $\mathcal{O}(1)$ coefficients, and don't
		contribute much to the other electroweak parameters.
		% We perform a global fit  of the various SM observables to the RS model parameters by constructing a $\chi^2$ statistic which is composed of:\newline
		%1)~A set of precisely measured observables\footnote{The notation is self explanatory} $\{m_Z,m_H,G_F,\alpha(m_Z),\alpha_s(m_Z),m_t\}$ is used as the input observables and taken to be free in the fit and are varied within their 1 standard deviation about their central value.\newline
		%2)~The output observables given by the set:
		%\begin{equation}
		%	\hat O _{k'}\equiv\{m_W,\Gamma_Z,\sigma_{had},R_e,R_\mu,R_\tau,R_b,R_c,\sin^2\theta_e,\sin^2\theta_b,\sin^2\theta_c,A^e_{FB},A^b_{FB},A^c_{FB},A_b,A_c \}
		%	\label{outputobs}
		%\end{equation}
		%The output observables are extracted for the value of the input observables which minimise the $\chi^2$ statistic. The relation between the input and the output observables is made explicit in the expansion formalism of \cite{Wells} and will be used for our analysis
		%New physics effects are introduced in the lagrangian in the form of higher dimensional operators which may directly contribute to the input observables in addition to the other precision observables which are labelled as the output observables. 
		%Owing to the precise determination of the input observables, the expression for the input observables must be redefined so that they lie within the experimentally allowed deviations about their central values. This redefinition will then affect the calculation of the output observables.  A $\chi^2$ statistic is then minimized using not only the model parameters but also the input observables. 
		We focus our attention on the scenarios a) and b) of the extensions to the RS model.
		In both the scenarios the brane mass parameter of the Higgs doublet plays an important role in determining constraint on the first KK mass of the gauge boson. Additionally with the precise
		measurement of the Higgs mass, the extent of fine tuning is related to the brane mass parameter $b$\cite{Cabrer1,Cabrer2,Cabrer3}. We make explicit the interplay between the fine tuning required to fit
		the Higgs boson mass and the $b$ parameter which gives the best fit to the electroweak observables.
		Using the expansion formalism of \cite{Wells} we determine the best fit points for the model parameters by using the Standard $\chi^2$ analysis. We find that the KK mass of the first gauge boson is lower than what was obtained in earlier analysis when naive bounds from evaluation of \textit{S} and \textit{T} parameters were taken into account. 
		
		The paper is organized as follows:
		In Section[\ref{section2}] and Section[\ref{section2a}] we outline the formalism of \cite{Wells} thus providing the necessary background for the analysis. in section[\ref{section3}] we briefly review the bulk RS model and study the bulk RS model with no additional gauge symmetries. In Section[\ref{section4}] and Section[\ref{section5}] we analyse RS model with deformed metric and custodial symmetry respectively. In Section[\ref{section6}] we conclude.

		\section{Expansion formalism and the SM}
		\label{section2}
		In this section we briefly review the expansion formalism of \cite{Wells} which we use for our analysis. There are numerous observables in the Standard Model whose values have
		been very well measured. These observables are in general a function of the following lagrangian parameters:
		\begin{equation}
			p_{k'}\equiv  \{g_i,y_t,v,\lambda\}
		\end{equation}
		where $g_i$ are the gauge couplings, $y_t$ is top quark Yukawa coupling, $v$ is vacuum expectation value (vev) and $\lambda$ is the quartic coupling. These parameters are referred to
		as the `input parameters'. An $i^{th}$ observable, $ \hat O_i^{SM}$ in the SM can be expressed as a function of these parameters as
		\begin{equation}
			\hat O_i^{SM}(\{p_{k'}\})=\hat O_i^{ref}+\sum_{k'}\frac{\partial\hat O_i^{SM}}{\partial p_{k'}}(p_{k'}-p_{k'}^{ref})+\ldots
			\label{smexp}
		\end{equation}
		where the $\ldots$ denote higher orders. $\{p_{k'}^{ref}\}$ is the set of lagrangian parameters chosen at a reference value $p^{ref}_{k'}$ at which the evaluated expressions for the SM observables match closely with experiment
		and $\hat O_i^{ref}=\hat O_i^{SM}(\{p_{k'}^{ref}\})$. Thus the expansion in Eq.(\ref{smexp}) is about the reference values ${p_{k'}^{ref}}$ and ${p_{k'}}$ is the allowed deviation about
		the reference values.
		
		The Lagrangian parameters, however are not measured directly, but are extracted from the measurements of certain observables. As a result it seems logical to re-express the SM observables
		in terms of a few accurately measured observables which will now serve as the input. One such list of input observables is\footnote{A subset of these observables can be used to `determine' the input parameters.}
		\begin{equation}
			\hat O_{k'}\equiv \{m_Z,m_H,G_F,\alpha(m_Z),\alpha_s(m_Z),m_t(m_t)\} 
		\end{equation}
		In terms  of the input observables, Eq.(\ref{smexp}) can be re-expressed as follows:
		\begin{equation}
			\hat O_i^{SM}(\{\hat O_{k'}\})=\hat O_i^{ref}+\sum_{k'}\frac{\partial\hat O_i^{SM}}{\partial \hat O_{k'}}(\hat O_{k'}-\hat O_{k'}^{ref})+\ldots
			\label{input}
		\end{equation}
		where $\hat O_{k'}^{ref}$ is the experimentally measured central value of the input observable and $\hat O_{k'}$ quantifies the deviation from the central value.
		Thus the deviation in $\hat O_{i}^{SM}$ can be expressed in terms of experimental
		deviation of the input observables from their central values. The relative deviation can be defined as
		\begin{equation}
			\bar\delta^{SM}\hat O_{i}(\{\hat O_{k'}\})=\frac{\hat O_i^{SM}(\{\hat O_{k'}\})-\hat O_i^{ref}(\{\hat O_{k'}^{ref}\})}{\hat O_{i}^{ref}}
			\label{deviation}
		\end{equation}
		Defining 
		\begin{equation}
			c_{ik'}=\frac{\hat O_{k'}^{ref}}{\hat O_i^{ref}}\frac{\partial \hat O_i^{SM}}{\partial\hat{ O}_{k'}}
			\label{corrcoeff}
		\end{equation}
		we can express the relative deviation in Eq.(\ref{deviation}) in the $i^{th}$ observable, due to deviation from the central value of the input observables as
		\begin{equation}
			\bar\delta O^{SM}_i=\sum_{i'}c_{ii'}\delta\hat O^{SM}_{i'}
			\label{deviation2}
		\end{equation}
		Using this, Eq.(\ref{input}) can be re-written as
		\begin{equation}
			\hat O_i^{SM}(\{\hat O_{k'}\})=\hat O_{i}^{ref}(1+\hat\delta^{SM}\hat O_i)
		\end{equation}

		The deviation of all the SM observables can be quantified by constructing a $\chi^2$ statistic   defined as 
		\begin{equation}
			\chi^2(\hat O_{k'})=\sum_i\left[\frac{\hat O_i^{SM}(\hat O_{k'})-\hat O_i^{expt}}{\delta \hat O_i^{expt}}\right]^2
			\label{chisq}
		\end{equation}
		It should be noted that while constructing the $\chi^2$ we assume that there is no correlation \cite{Falkowski:2014tna} between the output observables.
		However we note that taking into account correlation matrix for the output observables will not significantly change the results of our analysis.
		The central values and the allowed deviation for the input and output observables are given in Table[\ref{inout}]. Using the Z pole observables and the W mass as the output observables,
		we minimize the $\chi^2$ function in Eq.(\ref{chisq}) by varying the input observables within the experimentally allowed deviation given in Table(\ref{inout}). 
		The minimization is performed using \cite{minuit}.
		We obtain 
		$\chi^2_{min}=24.54$ with the corresponding best fit values for the input observables given in Table[\ref{smout}].
		
		\begin{table}[htbp]
			\begin{center}
				\begin{tabular}{|c|c|cc||c|cc|}
					\hline
					\multirow{3}{*}{Input observables} & $m_Z$ &91.1876(21)&\cite{ALEPH:2005ab}& $G_F$&$1.1663787(6)\times 10^{-5}$& \cite{Beringer:1900zz} \\
					&$\alpha(m_Z)$&$7.81592(86)\times10^{-3}$&\cite{Beringer:1900zz}&$m_t(m_t)$&173.34(75)&\cite{ATLAS:2014wva}\\
					&$\alpha_s(m_Z)$&0.1185(6)&\cite{Beringer:1900zz}&$m_H$&125.9(4)&\cite{Beringer:1900zz}\\
					\hline
					\hline
					\multirow{3}{*}{Output observables} &$m_W$&80.385(15)&\cite{Group:2012gb}&$\Gamma_Z$&2.4952(23)&\cite{ALEPH:2005ab}\\
					
					&$\sigma_{had}$&41.541(37)&\cite{ALEPH:2005ab}&$R_e$&20.804(50)&\cite{ALEPH:2005ab}\\
					&$R_{\mu}$&20.785(33)&\cite{ALEPH:2005ab}&$R_{tau}$&20.764(45)&\cite{ALEPH:2005ab}\\
					&$R_b$&0.21629(66)&\cite{ALEPH:2005ab}&$R_c$&0.1721(30)&\cite{ALEPH:2005ab}\\
					&$sin^2\theta_e$& 0.23153(16)&\cite{ALEPH:2005ab}&$sin^2\theta_b$& 0.281(16)&\cite{ALEPH:2005ab}\\
					&$sin^2\theta_c$& 0.2355(59)&\cite{ALEPH:2005ab}&$A^e_{FB}$& 0.0145(25)&\cite{ALEPH:2005ab}\\
					&$A^b_{FB}$&0.0992(16)&\cite{ALEPH:2005ab}&$A^c_{FB}$& 0.0707(35)&\cite{ALEPH:2005ab}\\
					&$A_b$&0.923(20)&\cite{ALEPH:2005ab}&$A_c$&0.670(27)&\cite{ALEPH:2005ab}\\  
					\hline
				\end{tabular}
				
			\end{center}
			\caption{Experimentally measured central values for the input and output observables along with the Standard Deviation}
			\label{inout}
		\end{table}

		\begin{table}[htbp]
			\begin{center}
				\begin{tabular}{|c|c|c||c|c|}
					\hline
					\multirow{3}{*}{Input observables} & $m_Z$ &91.188& $G_F$&$1.16638\times 10^{-5}$ \\
					&$\alpha(m_Z)$&$7.81589\times10^{-3}$&$m_t(m_t)$&173.59\\
					&$\alpha_s(m_Z)$&0.118567&$m_H$&125.89\\
					\hline
					\hline
					\multirow{3}{*}{Output observables} &$m_W$&80.366 $\pm0.005$&$\Gamma_Z$&2.4957$\pm0.0006$\\
					
					&$\sigma_{had}$ & 41.472 $\pm0.04$&$R_e$&20.7427$\pm0.03$\\
					&$R_{\mu}$&20.7428$\pm0.03$&$R_{tau}$&20.7897 $\pm0.03$\\
					&$R_b$&0.215822 $\pm0.00005$&$R_c$&0.17209$\pm0.000007$\\
					&${\rm sin}^2\theta_e$& 0.23161 $\pm0.000002$&${\rm sin}^2\theta_b$& 0.2329$\pm0.00001$\\
					&${\rm sin}^2\theta_c$& 0.2315$\pm0.000002$&$A^e_{FB}$& 0.0160$\pm0.000004$\\
					&$A^b_{FB}$&0.1025$\pm0.00002$&$A^c_{FB}$& 0.0732$\pm0.00001$\\
					&$A_b$&0.9346$\pm0.00005$&$A_c$&0.6675$\pm0.000008$\\  
					\hline
				\end{tabular}
				\label{smout}
			\end{center}
			\caption{Best fit values for the input and output observables for the SM fit with $\chi^2_{min}=24.54$}
		\end{table}

		\section{New Physics}
		\label{section2a}
		The expansion formalism presented in Section[\ref{section2}] can be extended to include 
		new physics effects. 
		Assuming the nature of new physics is such that it modifies mostly the oblique parameters,
		the new physics effects can be parametrized by the introduction of higher dimension operators in the lagrangian. These operators can give corrections to any of the input and the output
		observables. \footnote{In the calculations used for our analysis only tree level effective theory operators are considered. Things could get more stringent if one loop effective theory operators are considered \cite{Alonso:2013hga,Elias-Miro:2013eta,Henning:2014wua}. For a detailed analysis  of precision observables using Standard Model effective field theory see \cite{Trott}.}
		In the presence of new physics Eq. (\ref{deviation2}) becomes
		\begin{equation}
			\delta^{NP}\hat O_i^{th}(\{\hat O_{k'}\},NP)=\bar\delta \hat O^{SM}_i+\zeta_i
			\label{newphysics}
		\end{equation}
		where $\zeta_i(\{\hat O_{k'}\},NP)\equiv \frac{\delta^{NP}\hat O_i^{th}(\{\hat O_{k'}\},NP)}{\hat O_i^{ref}}$ parametrizes the relative contribution to the $i^{th}$
		observable due to higher dimension operators. Using Eq.(\ref{deviation2}), Eq.(\ref{newphysics}) can be written as 
		\begin{eqnarray}
			\delta^{NP}\hat O_i^{th}(\{\hat O_{k'}\},NP)&=&\sum_{i'}c_{ii'}\delta\hat O^{SM}_{i'}+\zeta_i\nonumber\\
			&=&\sum_{i'}c_{ii'}\delta\hat O^{th}_{i'}+\bar\delta^{NP}\hat O_i
			\label{newphysics2}
		\end{eqnarray}
		where $\bar\delta^{NP}\hat O_i=\zeta_i-\sum_{i'}c_{ii'}\zeta_{i'}$. 
		Here the superscript \textit{'th'} denotes SM in addition to new physics.
		Note that from Eq.(\ref{corrcoeff}), the matrix of co-efficients $c_{ik'}$ is a unit matrix for the input observables 
		\textit{i.e.} $c_{i'k'}=\delta_{i'k'}$. Thus any new physics effects to the input observables are adjusted such that the net shift is zero, which is apparent in Eq.(\ref{newphysics2}).
		This adjustment is however propagated in the evaluation of the output observables through Eq.(\ref{newphysics2}).
		
		In many scenarios, new physics is such that their dominant contribution to the various SM observables 
		is only through the self energy corrections to the various gauge boson propagators given below: 
		\begin{equation}
			\pi_{\alpha\beta}\equiv\{\pi_{ZZ},\pi'_{ZZ},\pi_{\gamma Z},\pi'_{\gamma\gamma},\pi_{WW},\pi^0_{WW}\} 
			\label{propagators}
		\end{equation}
		The primed quantities denotes differentiation with respect to $q^2$, where $q$ is the four momentum. 
		Note that the corrections to the fermion coupling to the gauge bosons are universal.
		In this case the new physics contribution to the input observables in Eq.(\ref{newphysics2})
		can be re-expressed as
		\begin{equation}
			\delta^{NP}\hat O_i^{th}=\sum b_{i,\alpha\beta}\delta^{NP}\pi_{\alpha\beta}
			\label{propexp}
		\end{equation}
		where it is understood that the sum extends over the list in Eq.(\ref{propagators}) while the coefficients $b_{\alpha\beta}$ are evaluated in \cite{Wells}. 
		
\begin{figure}[here]
			\centering
			\includegraphics[width=9cm]{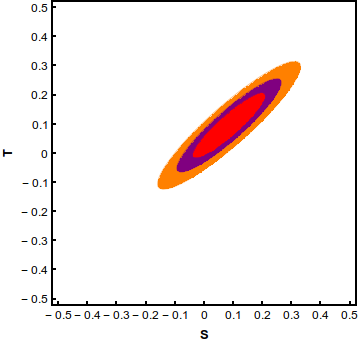}
			\caption{The red,blue and orange regions denote the 68\%,95\% and 99\% confidence level allowed regions in the ST parameter space.}
			\label{STplane}
		\end{figure}

		In such models the corrections to the gauge boson propagators can be encoded in oblique parameters \textit{S} and \textit{T} \footnote{The contribution to \textit{U} is suppressed as only dimension 8 operators contribute to it} \cite{Peskin:1990zt,Peskin:1991sw}.
		These oblique parameters are related to the new physics effects to the self energy correction as follows \cite{Ellis};
		\begin{eqnarray}
			\delta^{NP}\pi_{ZZ}&=&-\alpha(m_Z)T+\frac{\alpha(m_Z)}{2}S\nonumber\\\nonumber
			\delta^{NP}\pi'_{ZZ}&=&\frac{\alpha(m_Z)}{2}S\\\nonumber
			\delta^{NP}\pi_{\gamma Z}&=&-\frac{\alpha(m_Z)}{4{\rm sin}^2\theta_W}{\rm cos }2\theta_W{\rm tan}\theta_W S\\\nonumber
			\delta^{NP}\pi'_{\gamma \gamma}&=&-\frac{\alpha(m_Z)}{2}S
			\label{selfenergy}
		\end{eqnarray}
		We use Eq.(\ref{propexp}) in Eq.(\ref{newphysics2}) to construct the $\chi^2$ for the output observables at the Z peak along with the $W$ mass.
		Using the results of the analysis in \cite{Wells}, the expression for the $\chi^2$ statistic defined in Eq.(\ref{chisq}) is given as
		\begin{equation}
			\chi^2=25.0898+1102.39~S^2 +28.746~S-72.0085~T-2256.69~ST+1377.07~T^2
		\end{equation}
		The input observables were fixed to their experimentally measured central values while obtaining the above expression. 
		Using this we obtain the S-T plot in 
		Fig.[\ref{STplane}] in which the 68\%,95\% and 99\% confidence level allowed regions are depicted by red, blue and orange regions respectively. Our analysis is performed by fixing $U=0$.
		Additionally the best fit point for $m_h^{ref}=125$ GeV and $m_t^{ref}$ =173 GeV was obtained to be
		$S=0.08\pm 0.1$ and $T=0.09\pm 0.1$ with the correlation coefficient between the $S$ and the $T$ parameter to be $+0.89$.
		In addition to $S$ and $T$, the other floating point parameters were $m_Z=91.1876\pm 0.0021$, $\alpha_s(M_Z)=0.1185\pm 0.0006$ and $\alpha(M_Z^2)=7.81596(86)\times 10^{-3}$
		These results are to be compared with the Gfitter analysis \cite{Baak:2012kk}
		where the $\{m_Z,\alpha,\Delta\alpha\}$ were considered as the floating point parameters with $\Delta \alpha(M_Z^2)=0.02757\pm 0.0001$. They obtained $S=0.06\pm 0.09$ and $T=0.1\pm 0.07$ with a correlation co-efficient $+0.91$.

		%   The oblique parameters can be computed by evaluating the co-efficiencts of the following two higher dimensional operators
		% 
		% \begin{equation}
		%  \mathcal{O}_T=\frac{\alpha_T}{\Lambda^2}|HD_\mu H|^2\;\;\;;\;\;\;\;\mathcal{O}_{WB}=\frac{\alpha_s}{\Lambda}H^\dagger\tau^aHW^3_{\mu\nu}B_{\mu\nu}
		% \end{equation}
		% The co-efficients $\alpha_T,\alpha_S$ are related to the familiar oblique parameters $S,T$ \cite{} as follows
		% \begin{equation}
		%  S=\frac{8\pi v^2}{g_Lg_Y}\alpha_s\;\;\;\;\;;\;\;\;\;T=\frac{2\pi v^2}{e^2}\alpha_T
		% \end{equation}
		For a particular model of new physics, the oblique parameters \textit{S,T} depend on the model parameters. A given set of model parameters is valid only if the corresponding \textit{S,T} observables computed for that set lie at least within the orange ellipse in Fig.[\ref{STplane}].
		Thus a very small contribution, for example to the \textit{S} parameter would necessitate \textit{T} to also be very small so as to lie within the bottom left portion of the ellipse. However an increasing \textit{S} can admit larger values of the \textit{T} parameter corresponding to moving towards the top right portion of the ellipse.
		Thus we can use Fig.[\ref{STplane}] to constrain the model
		parameters. We now use this analysis to obtain constraints on various  Randall-Sundrum models.

		\section{Randall-Sundrum Models}
		\label{section3}
		Randall-Sundrum model is a model of a single extra-dimension compactified on an $S_1/Z_2$ orbifold \cite{RS}. The five dimensional gravity theory is defined by the following line element:
		\begin{eqnarray}
			ds^2=e^{-2A(y)}\eta_{\mu\nu}dx^\mu dx^\nu-dy^2
			\label{lineelement}
		\end{eqnarray}
		Two opposite tension branes are located at the two fixed points of the orbifold. The space between the branes is endowed with a large negative bulk cosmological constant making it 
		a slice of AdS. The presence of brane localized sources of energy results in zero cosmological constant being induced on the branes. In the original setup $A(y)=ky$ where k
		is reduced Planck scale. Identifying the scale of physics on the $y=0$ brane as $ M_{IR}$, the effective UV scale induced at the $y=\pi R$ brane owing to geometry is given 
		as
		\begin{equation}
			M_{IR}=e^{-kR\pi}k
		\end{equation}
		where $R$ is the compactification radius. Choosing $kR\sim 12$,  will result in $M_{IR}\sim 200$ GeV owing to large exponential warping. Any radiative instability to the masses of fundamental
		scalars in the theory can be warped down to the electroweak scale thus solving the gauge hierarchy problem. In the original setup, with the exception of gravity all the SM fields were
		localized on the brane at $y=y_1=\pi R$ also referred to as the IR brane. Here we consider a generalization of the original setup where particles of all types of spin are allowed to propagate in the bulk.
		
		A bulk field $\Psi^s(x^\mu,y)$ with spin $s$ can be expanded in the KK basis as follows:
		\begin{equation}
			\Psi^s(x^\mu,y)=\frac{1}{\sqrt{\pi R}}\sum^\infty_{n=0}\psi^{(n)}_s(x^\mu)f_s^{(n)}(y)
		\end{equation}
		The zero modes for the fields are identified as the SM fields. While the zero mode for the gauge bosons are flat at leading order, the ones for the scalars and the fermions are controlled
		by the brane and bulk mass terms respectively. They are parametrized as $m^s_{brane}=bk$ and $m^f_{bulk}=ck$ where $b,c$ are dimensionless $\mathcal{O}$(1) quantities. The normalized profiles 
		for the fields are given as
		\begin{eqnarray}
			f^{(0)}_0(b,y)&=&\sqrt{\frac{2(b-1)kR\pi}{e^{2(b-1)kR\pi}-1}}e^{(b-1)ky}\nonumber\\
			f^{(0)}_{1/2}(c,y)&=&\sqrt{\frac{(1-2c)kR\pi}{e^{(1-2c)kR\pi}-1}}e^{(0.5-c)ky}\nonumber\\
			f^{(0)}_1(y)&=&1
			\label{bulkprofiles}
		\end{eqnarray}
		where the normalization conditions are given as
		\begin{equation}
			\frac{1}{\pi R}\int_0^{\pi R}(f^{(0)}_s(y))^2dy=1
		\end{equation}
		$c<0.5$ and $b>1$ ($c>0.5$ and $b<1$ ) correspond to the fields being localized towards the IR(UV) brane respectively. The KK modes of all fields are however localized near the IR brane.
		We note here that while the profiles of the gauge boson fields are flat at leading order, it receives corrections due to the mixing of the KK mode with the zero mode. The mixing is proportional to the 
		vacuum expectation value (vev) and is given as
		\begin{equation}
			a_{01}=\frac{v^2}{(kR\pi)M^2_{KK}}\int_1^{z_{IR}}z^2f^{(0)}_0(z)^2f_1(z)^{(0)}f_1(z)^{(1)} 
			\label{mixing}
		\end{equation}
		where $z=e^{k R y}$ is the conformal co-ordinate. $f^{(0)}_{0,1}$ are given in Eq.(\ref{bulkprofiles}), while $f^{(1)}_{1}$ is the profile of the first KK mode of the gauge boson. Detailed review about bulk RS models can be found in \cite{Gher2,gherghetta}.
		Thus diagonalizing the mass matrix of KK modes and zero mode, will result in the lightest state (identified as the SM boson) having a small KK component proportional to
		Eq.(\ref{mixing}). As a result the coupling of the fermions to the SM boson will have a non-universal component which is a function of its localization parameter $c$.  The $c$ parameters will be in general different for different fermionic generations to generate the required hierarchy in the Yukawa parameters. 
		For the light fields with the exception of the top it is fair to assume $c>0.5$ to reduce the overlap with the Higgs. 
		For $c>0.5$, non-universal component of the coupling is very small and can be neglected \cite{Gher2,Hewett:2002fe}. This is enough to evade bounds from FCNC processes which can occur
		at tree level.
		
\begin{figure}
			\centering
			\includegraphics[width=9cm]{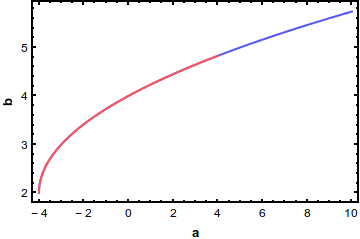}
			\caption{Maximum allowed value of $b$ for $\frac{k}{M_{pl}}=0.25$ (red) and $\frac{k}{M_{pl}}=0.1$(blue)  }
			\label{bulkscalar}
		\end{figure}

		Solution of the hierarchy problem requires the Higgs zero mode to be localized very close to the IR brane. It corresponds to a choice $b\geq2$ for the brane mass parameter\cite{Luty:2004ye,Cabrer2}\footnote{Realization
			of EWSB also requires $b>2$\cite{Davoudiasl:2005uu}.}.
		For a bulk scalar field with a massless zero mode the brane mass parameter $b$ is related to the bulk mass parameter $a$ as \footnote{Bulk scalar mass is parametrized as $m^s_{bulk}=ak$} 
		
		\begin{equation}
			b=2\pm\sqrt{4+a}
		\end{equation}
		Henceforth, we will drop the $b=2-\sqrt{4+a}$ solution as it will never lead to $b\geq2$ necessary for the solution to the hierarchy problem. The zero mode increasingly becomes sharply localized near the IR brane as $b$ increases. However an increase in $b$ is only facilitated by the corresponding increase in $|a|$. Depending on the value of $\frac{k}{M_{pl}}$, the bulk mass parameter cannot be increased indefinitely as the product $ak$ will become greater than the 5D Planck scale. Fig.[\ref{bulkscalar}] shows a plot of $b$ as function of $a$. Depending on the value of $\frac{k}{M_{pl}}$, the plot is terminated on the right at which $ak=M_{Pl}$. For instance for $\frac{k}{M_{pl}}=0.1$, the plot (blue curve) is terminated at $b=5.74$ while for $\frac{k}{M_{pl}}=0.25$ the plot (red curve) is terminated  $b=4.82$.
		
		%This can be seen by considering the following bulk scalar mass term with bulk mass $\propto k$ given as
		%\begin{equation}
		%\mathcal{L}^4_{mass}\in \int dy (f^{(0)}_0(b,y))^2 k ^2
		%\end{equation}  
		Thus the case with brane localized Higgs will be treated separately and not as a limiting case where the bulk Higgs field tends to a brane localized one.
		
		We now proceed to study the impact of EWPT in various RS models.
		\subsection{Bulk Higgs with no additional symmetries}
		This model is the same same set-up discussed above. All the fermionic fields except the top are localized near the UV brane. This 
		is sufficient 
		to fit the masses of all fermions except the top. 
		Due to localization of all the fermions near the UV brane the vertex corrections are very small and universal, thus the new physics effects can be parametrized in terms of the oblique 
		operators \textit{S} and \textit{T} which are given as \cite{Cabrer1,Cabrer2,Cabrer3}
		\begin{eqnarray}
			\alpha T&=&\frac{{\rm sin}^2\theta_W m_Z^2 y_1 k^2e^{-2kR\pi}}{\Lambda^2_{IR}}(\alpha_{hh}-2\alpha_{hf}+\alpha_{ff})\nonumber\\
			\alpha S&=&\frac{8{\rm sin}^2\theta_W {\rm cos}^2\theta_W m_Z^2 y_1 k^2e^{-2kR\pi}}{\Lambda^2_{IR}}(\alpha_{ff}-\alpha_{hf})\nonumber\\
			\label{oblique}
		\end{eqnarray}
		where $y_1$ denotes the position of the IR brane and $\alpha_{ij}$ are parameters involving the bulk propagators of the bulk gauge fields with $(++)$ boundary conditions, where $+$ denotes Neumann boundary condition. They
		are given as \cite{Cabrer,Cabrer2}
		\begin{eqnarray}
			\alpha_{hh}&=&\int e^{2A(y)}\left(\Omega_h-\frac{y}{y_1}\right)^2\nonumber\\
			\alpha_{hf}&=&\int e^{2A(y)}\left(\Omega_h-\frac{y}{y_1}\right)\left(\Omega_f-\frac{y}{y_1}\right)\nonumber\\
			\alpha_{ff}&=&\int e^{2A(y)}\left(\Omega_f-\frac{y}{y_1}\right)^2
		\end{eqnarray}
		
		where $\Omega_{f,h}(y)=\frac{1}{y_1}\int_0^y dy f^2_{f,h}(y)$ and the profiles $f's$ are given by Eq.(\ref{bulkprofiles}) 
		%The subscript denotes that they are co-efficients of the following operators in the effective theory:
		%\begin{equation}
		% (j_\mu)_H(j^\mu)_H(\text{HH}) ; \;\;\;(j_\mu)_H(j^\mu)_F(\text{HF}); \;\;\; (j_\mu)_F(j^\mu)_F(\text{FF})  
		%\end{equation}
		% with $F,H$ denoting fermionic and the scalar current respectively
		For the case where the fermions are localized on the UV brane $\Omega_f=1$. 
		These co-efficients are a function of the localization of the zero mode of the fermionic and the Higgs field. 
		For a fixed KK scale, the co-efficients increase as the fields move closer
		to the IR brane due to larger overlap of the zero mode with the KK modes.
		
\begin{figure}[here]
			\centering
			\includegraphics[width=9cm]{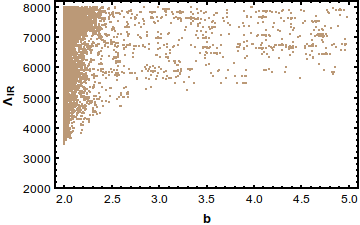}
			\caption{$3\sigma$ allowed parameter space in the $b-\Lambda_{IR}$ plane for regular bulk RS. $\Lambda_{IR}$ is in GeV.}
			\label{bmkknormal}
		\end{figure}

		For the oblique $T$ parameters, the co-efficient  $\alpha_{hh}$, also contributes in addition to $\alpha_{hf,ff}$. Owing to the localization of the Higgs very
		close to the IR brane, $\alpha_{hh}$ will be enhanced as compared to $\alpha_{hf,ff}$, which is smaller as the fermions are closer to the UV brane. As a result
		in this scenario the contributions to the $T$ parameter is large. In this case the oblique observables primarily depend on two parameters: \newline
		\textbf{a)}~The localization parameter $b$ for the bulk Higgs field.\newline
		\textbf{b)}~First KK scale of the gauge boson.\newline
		To extract the parameter space of these two parameters which are consistent with the constraints on the \text{S} and \textit{T} parameters, a scan is performed over the following ranges
		\begin{equation}
			b\equiv[2,5]\;\;\;\;\;\;\;\;\;\;\;\;\;\;\Lambda_{IR}\equiv[1250,10000]
		\end{equation}
		Fig.[\ref{bmkknormal}] shows the $3\sigma$ region in the $b-\Lambda_{IR}$ plane. 
		The first KK mass of the
		gauge boson is related to the IR scale as $m^{(1)}_{KK}\sim 2.44\Lambda_{IR}$. We see $\Lambda_{IR}$ is lowered as $b$ approaches 2 corresponding to shifting of the Higgs away from 
		the IR brane. However $b\geq 2$ must be maintained for the model to serve as solution to the hierarchy problem \cite{Luty:2004ye,Cabrer2}. 
		%The lowest KK scale allowed for $b=2$ is $6.56$ TeV which is clearly
		%outside the reach of LHC. 
		%In this case the input observables are fixed to their experimentally measured central values.
		Table[\ref{normalRSout}] gives the  fit values
		when all input observables along with $b$ and $\Lambda_{IR}$ are varied simultaneously to minimize the $\chi^2$ in Eq.(\ref{chisq}). From the plot in
		Fig.[\ref{bmkknormal}] we find that the lowest value of $\Lambda_{IR}$ possible is around 3.4 TeV corresponding a first KK mass for the gauge boson to be around 8 TeV. The plot is highly concentrated around $b=2$ since the coupling of the SM fields to the KK states is small as compared to higher values of $b$.
		This point corresponds to the case where the 
		mass of the first KK gauge boson is minimum. 
		%Varying the input observables has the effect of lowering the KK scale to about 5.9 TeV at 3$\sigma$  as shown in Fig.[\ref{bmkknormal}].

		\begin{table}[htbp]
			\begin{center}
				\begin{tabular}{|c|c|c||c|c|}
					\hline
					\multirow{3}{*}{Input observables} & $m_Z$ &91.1813& $G_F$&$1.1663784\times 10^{-5}$ \\
					&$\alpha(m_Z)$&$7.81611\times10^{-3}$&$m_t(m_t)$&173.05\\
					&$\alpha_s(m_Z)$&0.119101&$m_H$&126.3\\
					\hline
					\hline
					\multirow{3}{*}{Output observables} & $m_W$&80.411&$\Gamma_Z$& 2.4983\\
					&$\sigma_{had}$&41.479&$R_e$&20.7472\\
					&$R_{\mu}$&20.7473&$R_{tau}$&20.7941\\
					&$R_b$&0.2158&$R_c$& 0.1712\\
					&${\rm sin}^2\theta_e$&0.2313&${\rm sin}^2\theta_b$ &0.2327\\
					&${\rm sin}^2\theta_c$&0.2312&$A^e_{FB}$ & 0.0164\\
					&$A^b_{FB}$&0.1038&$A^c_{FB}$ & 0.0742\\
					&$A_b$&0.9347&$A_c$&0.6683\\ 
					\hline
					Model Parameters		& b &2.00& $m^1_{kk}$& 8.3 TeV\\   
					\hline 
					\hline
				\end{tabular}
				
			\end{center}
			\caption{Fit values for the input and output observables with a bulk Higgs. Input observables are free in the fit and are varied within their 1 $\sigma$ allowed experimental deviation. b=2.00 and 
				$m^1_{kk}=8.3$ TeV is obtained for the fit.}
			\label{normalRSout}
		\end{table}
		We finally note that for the brane localized case, a minimum KK mass of 13.6 TeV is required 
		for the model to be consistent with the data. The fit values for the input and the output observables are given in Table[\ref{normalRSoutbrane1}]
		
		\begin{table}[htbp]
			\begin{center}
				\begin{tabular}{|c|c|c||c|c|}
					\hline
					\multirow{3}{*}{Input observables} & $m_Z$ &91.1856& $G_F$&$1.16637854\times 10^{-5}$ \\
					&$\alpha(m_Z)$&$7.816649\times10^{-3}$&$m_t(m_t)$&172.33\\
					&$\alpha_s(m_Z)$&0.118657&$m_H$&126.295\\
					\hline
					\hline
					\multirow{3}{*}{Output observables} & $m_W$&80.402&$\Gamma_Z$& 2.4976\\
					&$\sigma_{had}$&41.477&$R_e$&20.744\\
					&$R_{\mu}$&20.744&$R_{tau}$&20.7916\\
					&$R_b$&0.2158&$R_c$& 0.17229\\
					&${\rm sin}^2\theta_e$&0.2314&${\rm sin}^2\theta_b$ &0.2327\\
					&${\rm sin}^2\theta_c$&0.2313&$A^e_{FB}$ & 0.0164\\
					&$A^b_{FB}$&0.1036&$A^c_{FB}$ & 0.07412\\
					&$A_b$&0.9347&$A_c$&0.6682\\  
					\hline
				\end{tabular}
			\end{center}
			
			\caption{Fit values for the input and output observables. Input observables are free in the fit and are varied within their 1 $\sigma$ allowed experimental deviation. 
				$m^1_{kk}=13.6$ TeV is obtained for the fit.}
			\label{normalRSoutbrane1}
		\end{table}
		
		It is to be noted that KK scales in excess on 20 TeV
		is required when constraints from FCNC like $\mu\rightarrow e\gamma$ are taken into account \cite{Iyer:2012db}. Implementation of bulk flavour symmetries with the imposition of the Minimal Flavour Violation (MFV) ansatz helps in substantially reducing the KK mass to around $\sim 3$ TeV \cite{AgasheSoni,perez,Fitzpatrick,Iyer} 
		
		% \begin{table}[htbp]
		%  \begin{center}
		%   \begin{tabular}{|c|c|c|c|c|c|c|}
		%    \hline
		%    Input %Observable&$m_Z$(Gev)&$G_F$(GeV$^{-2}$)&$\alpha(m_Z)$&$m_t$(Gev)&$\alpha_s(m_Z)$&$m_H$(GeV)\\
		%    \hline
		%    Best Fit&91.1813&1.166379242e$^{-5}$&0.00781668553&173.05& 0.11852&126.299\\
		%    \hline
		%   \end{tabular}
		%   \label{normalRS}
		%  \end{center}
		%  \caption{Best fit values for the input parameters for the normal RS case with %$\chi^2_{min}=23.01$. The corresponding RS parameters are $b=2.00$ and $m^1_{kk}=9.45$ TeV}
		% \end{table}

		Fine Tuning: It is well known that the requirement of massless zero mode for bulk scalar field requires the bulk mass $m_{bulk}=ak$ and the 
		brane mass $m_{brane}=bk$  be related by the 
		following relation\footnote{The relation $m_{brane}=2k-\sqrt{4k^2+m^2_{bulk}}$ is relevant for a Higgs field localized away from the IR brane and is not relevant to
			the discussion here.}:
		\begin{equation}
			m_{brane}=2k+\sqrt{4k^2+m^2_{bulk}}
		\end{equation}
		Any misalignment between the brane and the bulk masses will result in a non-zero mass for the zero mode. In a realistic model with electroweak symmetry breaking, the Higgs boson is massive
		and is related to the misalignment as\cite{Cabrer1,Cabrer2,Quiros:2013yaa}:
		\begin{equation}
			m_H^2=4(b-1)(m_{brane}-m'_{brane})\frac{\Lambda^2_{IR}}{k}
			\label{higgsmass}
		\end{equation}
		where $m'_{brane}=bk$ is value of the brane mass when the zero mode is massless. As given in Table[\ref{normalRSout}], $\Lambda_{IR}=2.41$ TeV for the model to consistent with the 
		electroweak precision data. As a result for $b=2$ a cancellation upto the fourth decimal between $m'_{brane}$ and $m_{brane}$ is required to fit  a Higgs mass of 126 GeV.
		The level of tuning increases as the Higgs field is pushed further near the IR brane corresponding to an increase in $b$. Due to a direct 
		dependence on the brane mass parameter $b$ \cite{Cabrer1,Cabrer2,Fitzpatrick:2013twa},
		it is fair to expect that the Higgs boson mass is best fit by $b\sim 2$. In the dual theory this corresponds to the Higgs field is a partial composite state with relevant coupling
		between the source and the CFT (conformal field theory) sectors. As $b$ increases, this coupling  increases and the state is fully composite of the CFT, thus recovering the original RS setup.

		%In order to alleviate the constraints on the KK scale from the precision test as well just to reduce the fine tuning to obtain the Higgs mass
		%the following two modifications of the RS models were considered:
		%a) RS model with deformed metric.\newline
		%b) RS model with a bulk custodial.\newline
		%c)Inclusion of brane kinetic terms.\newline
		%In this paper we will consider the first two possibilities.
		
		\section{Deformed RS model}
		\label{section4}
		The expression for the \textit{S} and \textit{T} parameters in Eq.(\ref{oblique}) can be re-expressed as \cite{Cabrer1,Cabrer2}
		\begin{eqnarray}
			\alpha S&=&8{\rm cos }^2\theta_W{\rm sin}^2\theta_W\frac{m_Z^2}{\Lambda^2_{IR}}\frac{1}{Z}I\nonumber\\
			\alpha T&=&{\rm sin}^2\theta_W\frac{m_Z^2}{\Lambda^2_{IR}}\frac{ky_1}{Z^2}I
			\label{oblique2}
		\end{eqnarray}
		where $y_1$ is the position of the IR brane and the dimensionless integral $I$ and $Z$ are defined as
		\begin{eqnarray}
			I=k\int_0^{y_1}\left[(k\left(y_1-y\right))^2\right]e^{2A(y)-2A(y_1)}dy\nonumber\\
			Z=k\int_0^{y_1}dy\frac{h^2(y)}{h^2(y_1)}e^{-2A(y)+2A(y_1)}
		\end{eqnarray}
		$h(y)$ is the profile of the vacuum expectation value and is given as
		\begin{equation}
			h(y)=h(y_1)e^{bk(y-y_1)}
		\end{equation}
		We find the $T$ parameter is enhanced by the volume factor in addition to being suppressed by two powers of $Z$.
		In RS models where $A(y)=ky$, $Z=0.5$ for $b=2$ and becomes smaller as the Higgs field approaches the IR brane ($b\rightarrow\infty$). This results in the enhancement of the 
		$T$ parameter leading to stringent constraints in the KK scale. As a result, the authors in \cite{Cabrer1,Cabrer2,Cabrer3} considered an alternative solution by considering modification of the line element
		in Eq.(\ref{lineelement}) where $A(y)$ is now given as
		\begin{equation}
			A(y)=ky-\frac{1}{\nu^2}\log(1-\frac{y}{y_s})
		\end{equation}
		Note that $\nu\rightarrow\infty$ results in RS limit. A consequence of this metric is that the singularity at the IR brane is shifted outside the patch between IR and UV brane
		at $y_s=y_1+\Delta$.  $\Delta$ is the distance of the singularity from the IR brane. 
		For the case where the hierarchy problem is solved \textit{i.e.} $A(y_1)\sim 36$, the position of the IR brane in the bulk $y_1$ is a function of $\nu,\Delta$.
		Smaller $\nu$ will in general result in a smaller volume factor $y_1$ and helps in ameliorating the constraints on the KK mass from the T parameter.
		Additionally as noted in \cite{Cabrer1,Cabrer2} this setup results in large values of $Z$ for certain choices of
		parameters $\nu,\Delta,b$ which help in reducing the KK scales so as to be within the reach of LHC. 
		
		As before we perform an analysis to determine the parameter space of the $b-\Lambda_{IR}$ plane. We choose two sets of $(\nu,\Delta)$ as follows:\newline
		\textbf{a)}~ $\nu=0.8$ and $\Delta=1$.This corresponds to $y_1=30.60/k$ so that $A(y_1)\equiv 36$\newline
		\textbf{b)}~ $\nu=1$ and $\Delta=0.1$.This corresponds to $y_1=30.28/k$ so that $A(y_1)\equiv 36$
		
		\begin{figure}[here]
			\begin{tabular}{cc}
				\includegraphics[width=9cm]{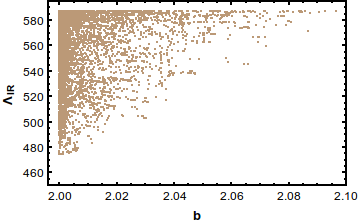}&\includegraphics[width=9cm]{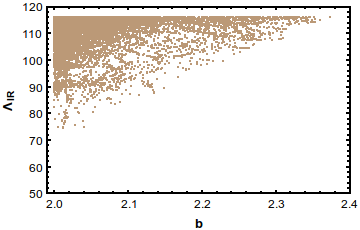} \\
			\end{tabular}
			\caption{Allowed parameter space in the $b-\Lambda_{IR}$ plane for deformed metric. $\Lambda_{IR}$ is in GeV. The left panel corresponds to $\nu=0.8$ and $\Delta=1$ while the right panel corresponds to $\nu=1$ and $\Delta=0.1$ }
			\label{bmkkdeformed}
		\end{figure}
		For the deformed metric, the $\Lambda_{IR}$ is related to the first $m_{kk}$ scale from the following relation\cite{Cabrer1,Cabrer2}
		\begin{equation}
			m^1_{kk} \sim j_{0,1}\frac{A'(y_1)}{k}\Lambda_{IR}
		\end{equation}
		where $j_{0,1}$ is the first zero of Bessel function $J_0(x)$. 
		We scan the $b$ parameter from 2 to 5 and $\Lambda_{IR}$ is scanned from 50 to 587 GeV for case a) while it is scanned from 50 to 120 GeV for case b).
		The upper limit on $\Lambda_{IR}$ corresponds to a KK mass of $\sim 3$ TeV.
		From Fig.(\ref{bmkkdeformed}), for the left panel, a lowest value of $\Lambda_{IR}=472.6$ GeV is obtained for b=2 which corresponds to a first KK mass of about 2.3 TeV.
		While for case b) depicted in the right panel of Fig.(\ref{bmkkdeformed}),
		a lowest value of $\Lambda_{IR}=71.10$ GeV is obtained again for
		$b=2$. This corresponds to a first KK mass of about 1.7 TeV for the gauge boson.
		Thus we see that for certain choices of the metric depending on the values $(\nu,\Delta)$,
		the first KK mass of the gauge boson can be below 2 TeV.
		The fit values for the input and output observables are given in Table[\ref{deformedRSout}].
		Case b) offers an advantage over Case a) in terms of being a less fine tuned model since the
		$\Lambda_{IR}$ for the fit is small.
		The analysis can be repeated for different values of $\nu$ and $\Delta$. For our analysis we fit the top quark mass by $c_{Q_3}\sim0.475$ and $c_{t}\sim -1$ with a choice of $\mathcal{O}(1)$ Yukawa $\sim 4$.
		
		The localization of the top doublet relatively near the UV brane is to minimize the correction to the $Zbb$ vertex.

		% \begin{table}[htbp]
		%  \begin{center}
		%   \begin{tabular}{|c|c|c|c|c|c|c|}
		%    \hline
		%    Input %Observable&$m_Z$(Gev)&$G_F$(GeV$^{-2}$)&$\alpha(m_Z)$&$m_t$(Gev)&$\alpha_s(m_Z)$&$m_H$(GeV)\\
		%    \hline
		%    Fit&91.1813&1.166378541e$^{-5}$&0.0078166312233&173.128& 0.11918&126.299\\
		%    \hline
		%   \end{tabular}
		%   \label{deformedRS1}
		%  \end{center}
		%  \caption{Fit values for the input parameters for the deformed RS case with %$\chi^2_{min}=32.97$. The corresponding RS parameters are $b=2.0006$ and $m^1_{kk}=1.68$ TeV.
		%  $\nu=0.1$ and $k\Delta=1$ are chosen for the fit}
		% \end{table}
		
		% The fit values for the output observables are given as
		\begin{table}[htbp]
			\begin{center}
				\begin{tabular}{|c|c|c||c|c|}
					\hline
					\multirow{3}{*}{Input observables} & $m_Z$ &91.1813& $G_F$&$1.1663785\times 10^{-5}$ \\
					&$\alpha(m_Z)$&$7.81663\times10^{-3}$&$m_t(m_t)$&173.12\\
					&$\alpha_s(m_Z)$&0.119118&$m_H$&126.29\\
					\hline
					\hline
					\multirow{3}{*}{Output observables} &$m_W$&80.419&$\Gamma_Z$&2.498\\
					&$\sigma_{had}$&41.486&$R_e$&20.7381\\
					&$R_{\mu}$&20.7382&$R_{tau}$&20.785\\
					&$R_b$&0.215&$R_c$&0.171\\
					&${\rm sin}^2\theta_e$&0.2314&${\rm sin}^2\theta_b$&0.2328\\
					&${\rm sin}^2\theta_c$&0.2313&$A^e_{FB}$&0.0162\\
					&$A^b_{FB}$&0.1032&$A^c_{FB}$&0.0737\\
					&$A_b$&0.9346&$A_c$&0.6679\\
					\hline
					Model Parameters		& b &2.00& $m^1_{kk}$& 2.3 TeV\\   
					\hline
				\end{tabular}
				\label{deformedRSout}
			\end{center}
			\caption{Fit values for the input and output observables for the deformed RS case. $b=2.0006$ and $m^1_{kk}=2.3$ TeV is obtained for the fit. $\nu=0.8$ and $\Delta=\frac{1}{k}$ are chosen for the fit}
		\end{table}
		
		%   \begin{table}[htbp]
		%   \begin{center}
		%    \begin{tabular}{|c|c|c|c|c|c|c|}
		%     \hline
		%     Input Observable&$m_Z$(Gev)&$G_F$(GeV$^{-2}$)&$\alpha(m_Z)$&$m_t$(Gev)&$\alpha_s(m_Z)$&$m_H$(GeV)\\
		%     \hline
		%     Best Fit&91.1813&1.166379175e$^{-5}$&0.00781607594&173.05& 0.118640&126.299\\
		%     \hline
		%    \end{tabular}
		%    \label{deformedRS2}
		%   \end{center}
		%   \caption{Best fit values for the input parameters for the normal RS case with $\chi^2_{min}=22.21$. The corresponding RS parameters are $b=2.17$ and $m^1_{kk}=0.511$ TeV.
		%   $\nu=1$ and $k\Delta=0.1$ are chosen for the fit}
		%  \end{table}
		Fine Tuning: Due to the deformation in the metric, the Higgs mass in Eq.(\ref{higgsmass}) can be generalized to \cite{Cabrer1,Cabrer2,Quiros:2013yaa}:
		\begin{equation}
			m_H^2=\frac{2}{Z}(m_{brane}-m'_{brane})\frac{\Lambda^2_{IR}}{k}\;\;\;\text{where}~Z=\int_0^{y_1}\frac{h^2(y)}{h^2(y_1)}e^{-2A(y)+2A(y_1)}
			\label{higgsmassdef}
		\end{equation}
		For the normal RS case $A(y)=ky$ and $Z=\frac{1}{2(a-1)}$, thus reducing to Eq.(\ref{higgsmass}). 
		In comparison to RS where $Z<1$, certain choices of $\nu$ and $\delta$ result in $Z>1$ which not only lowers the contribution to the T paramter in Eq.(\ref{oblique2}) but also
		helps in reducing the fine tuning to obtain the Higgs mass. For instance for the parameters in Table.[\ref{deformedRSout}], $Z=2.6$, the tuning reduces to 0.018. 
		
		\section{Custodial RS }
		\label{section5}
		The custodial Randall Sundrum set up [\cite{Agashe:2003zs}] contains an enlarged bulk gauged symmetry given by
		\begin{equation}
			SU(2)_L\times SU(2)_R\times U(1)_X
		\end{equation}
		which restores the custodial symmetry in the RS setup for the Higgs potential.
		The corresponding gauge bosons are denoted by $W^{1,2,3}_{L\mu},W^{1,2,3}_{R\mu},X_\mu$ 
		with $g_{5L,5R,5X}$ denoting the corresponding five dimensional gauge couplings. In updating the electroweak constraints in this setup we follow the notation of \cite{Carena:2006bn}.
		
		The bulk symmetry is broken down to the Standard Model by considering the following boundary conditions for the gauge fields
		\begin{equation}
			W^{1,2,3}_{L\mu}(++)\;\;\;B_{\mu}(++)\;\;\;W^{1,2}_{R\mu}(-+)\;\;\;Z'_{\mu}(-+)
		\end{equation}
		with $+(-)$ denoting Neumann(Dirichlet) boundary conditions as before.
		The gauge fields $B_{\mu}$ and $Z'_{\mu}$ are defined as
		\begin{equation}
			B_\mu=\frac{g_{5X}W^3_{R\mu}+g_{5R}X_{\mu}}{\sqrt{g^2_{5R}+g^2_{5X}}}\;\;\;\;\;\;;Z'_\mu=\frac{g_{5R}W^3_{R\mu}-g_{5X}X_{\mu}}{\sqrt{g^2_{5R}+g^2_{5X}}}
		\end{equation}
		The  the $W^{1,2,3}_{L\mu}$ and $B_\mu$ possess zero modes corresponding to the SM $SU(2)_L$ and the $U(1)_Y$ gauge
		boson respectively. The hypercharge coupling is given by $Y=2(T^3_R+Q_X)$.
		After electroweak symmetry breaking, the electromagnetic charge is given by $Q_{em}=T^3_R+T^3_L+Q_X$. 
		On the other hand, owing to the mixed boundary conditions of $W^{1,2}_{R\mu}$ and $Z'_{\mu}$, they do not
		possess a zero mode.
		
\begin{figure}
			\begin{center}
				\begin{tabular}{cc}
					\includegraphics[width=.5\textwidth]{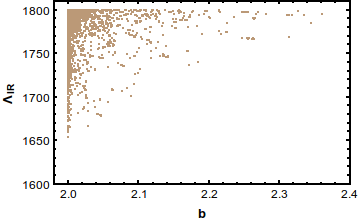}&\includegraphics[width=.5\textwidth]{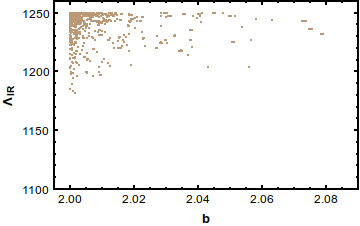}
				\end{tabular}
				\caption{Left panel shows the $b-\Lambda_{IR}$ parameter space when just the tree level computations of $S-T$ are taken into account. 
					In the right panel, the loop contributions to the $T$ parameter are also included. $\Lambda_{IR}$ is in GeV.}
				\label{custodial}
			\end{center}
		\end{figure}

		The presence of new gauge bosons induce additional corrections to the  \textit{T} parameter but the \textit{S} parameter remains unchanged. It is given as \cite{Cabrer2,shrihari,Delgado:2007ne,neubert1,Casagrande:2010si}
		\begin{equation}
			\alpha T=\frac{{\rm sin}^2\theta_W m_Z^2 y_1 k^2e^{-2kR\pi}}{\Lambda^2_{IR}}\left(\alpha_{hh}-\alpha'_{hh}-2\alpha_{hf}+\alpha_{ff}\right)
		\end{equation}
		where the $\alpha'_{hh}$ is the bulk propagator for the bosons with $(-+)$ boundary conditions and is given as
		\begin{equation}
			\alpha'_{hh}=\int_0^{y_1}e^{2A(y)}(1-\Omega_h)^2
		\end{equation}
		The contribution to  the $T$ parameter due to the KK states of the SM as well as the new gauge bosons are very similar in magnitude. Recalling that the dominant contribution
		to the $T$ parameter is due to $\alpha_{hh}$, the presence of $\alpha'_{hh}$ nearly
		cancels this contribution, thus significantly lowering the constraint on the first KK scale of the gauge boson.
		As a result, the dominant constraint to the KK scale is due to the $S$ parameter. To see the effects of a negligible T parameter on the fits we assume 
		UV localized fermions to begin with. To obtain the plot for $b-\Lambda_{IR}$ parameter space in the presence of custodial symmetry, we  first evaluate the constraints at tree level.
		% We note here that owing to the vanishing of the $T$ parameter at tree level, the dominant constraint on the KK mass comes from the evaluation of the $S$ parameter. 
		From the $S-T$ plot in Fig.[\ref{STplane}], the region around  $T\sim 0$ would also necessitate the $S$ parameter to be small thereby pushing the KK scale up. Indeed, as is noted in the left panel of Fig.[\ref{custodial}], a lowest value of $\Lambda_{IR}=1659$ GeV is obtained which translates into a lowest KK mass of around $4$ TeV. While this case does better than the normal bulk RS scenario the first KK mass is still out of reach of LHC.

		\begin{table}[htbp]
			\begin{center}
				\begin{tabular}{|c|c|c||c|c|}
					\hline
					\multirow{3}{*}{Input observables} & $m_Z$ &91.1938& $G_F$&$1.1663787\times 10^{-5}$ \\
					&$\alpha(m_Z)$&$7.81509\times10^{-3}$&$m_t(m_t)$&173.3499\\
					&$\alpha_s(m_Z)$&0.119003&$m_H$&125.40\\
					\hline
					\hline
					\multirow{3}{*}{Output observables} &$m_W$&80.347&$\Gamma_Z$&2.495\\
					&$\sigma_{had}$&41.471&$R_e$&20.736\\
					&$R_{\mu}$&20.736&$R_{tau}$&20.783\\
					&$R_b$&0.215&$R_c$&0.171\\
					&${\rm sin}^2\theta_e$&0.231&${\rm sin}^2\theta_b$&0.233\\
					&${\rm sin}^2\theta_c$&0.2317&$A^e_{FB}$&0.0155\\
					&$A^b_{FB}$&0.100&$A^c_{FB}$&0.072\\
					&$A_b$&0.934&$A_c$&0.666\\  
					\hline
					\hline
					Model Parameters		& b &2.00& $m^1_{kk}$& 2.88 TeV\\   
					\hline
				\end{tabular}
				\label{custodialRSout}
			\end{center}
			\caption{Fit values for the input and output observables. The corresponding RS parameters are $b=2.0004$ and $m^1_{kk}=2.9$. TeV. $\Lambda_{IR}$ is in GeV.
				The loop contribution to the $T$ parameter $\sim 0.06$}
		\end{table}
		
		The assumption of UV localized fermions is not sufficient to fit the top quark mass as it would  result in large $\mathcal{O}(1)$ Yukawa coupling. As a result the zero mode
		top doublet and the singlet must be moved closer to the IR brane ($c<0.5$) to increase overlap with the Higgs.  This results in the shift of the coupling of $b_L$ to the Z boson. The relative shift to the $b_L$ is given as \cite{ponton,Carena:2006bn,Delgado:2007ne}
		\begin{equation}
			\frac{\delta g_{b_L}}{g_{b_L}}=-v^2\frac{\left(g_L^2T^3_L-g_R^2T^3_R\right)\alpha'_{hf}+\left(g_L^2T^3_L-(g')^2Y\right)\left(\alpha_{hf}-\alpha^{UV}_{hf}-\alpha'_{hf}\right)}{1-\frac{2}{3}\sin^2\theta_W}
			\label{Zbb}
		\end{equation}
		The first term involving to $\alpha'_{hf}$ will be significant in this case as the third generation doublet is localized closer to the IR brane to fit the top quark mass.
		In the second term however the presence of $\alpha'_{hf}$ and $\alpha_{hf}$ with a relative minus sign softens the impact of localization of the third generation on $\frac{\delta g_{b_L}}{g_{b_L}}$.
		The current constraints on the corrections to the $Zb_Lb_L$ coupling
		%\begin{equation}
		% \frac{\delta g_{b_L}}{g_{b_L}}\leq
		%\end{equation}
		pushes the limit obtained in the left panel of Fig.[\ref{custodial}] to beyond 5 TeV. However it was observed in \cite{zbb}, that the dominant contribution due to
		the first term in Eq.(\ref{Zbb}), can be removed by assuming $T^3_L=T^3_R$ and $g_L=g_R$ thus significantly softening the constraints on the KK mass from corrections to the $Zb\bar b$ vertex. 
		This implies that the left handed bottom must belong to bi-doublets of $SU(2)_L\times SU(2)_R$. The bi-doublets induce large negative contributions to the T parameter in most regions of the 
		parameter space. The contribution is a function of $c_{Q_3,t}$ which are the localization parameters for the bi-doublet and the singlet $t_R$. It was noted in \cite{Carena:2006bn}
		that the negative contribution decreases as the doublet and/or singlet are localized away from the IR brane. However, 
		the top quarks mass as a function of $c_{Q_3,t}$ is given as
		\begin{equation}
			m_{top}=\tilde Y^u_{3,3}\frac{v}{\sqrt{2}(\pi R)^{3/2}}\int_0^{\pi R}f^{(0)}_0(b,y)f^{(0)}_{1/2}(c_{Q_3},y)f^{(0)}_{1/2}(c_{t},y)
		\end{equation}
		where $Y^u_{3,3}$ is dimensionless $\mathcal{O}$(1) parameter.
		Choosing the bi-doublet and the singlet to be localized away from the IR brane, will result in the choice of large $Y^u_{3,3}>10$ to fit the top quark mass and is not feasible.
		As s result the combination of $c_{Q_3,t}$ which induces a positive contribution to the T parameter is when $c_Q\leq 0$ and $c_t\sim[0.4,0.5]$. As shown by \cite{Carena:2006bn}, this choice of $c$ parameters not only fits the top quark mass but also gives non-negative contribution to the T parameter. 
		
		A non-zero positive contribution to the $T$ parameter would correspond to moving vertically up in the $S-T$ plane in Fig.[\ref{STplane}]. Owing the tilted orientation of the ellipse,
		it makes it possible to accommodate larger values of $S$ thereby helping in reducing the lower
		bound on the first KK mass. The right panel of Fig.[\ref{custodial}] corresponds to the case where the loop level contributions to the $T$ parameter \cite{Carena:2006bn} have been turned on. 
		In the figure we have assumed $\Delta T_{loop}\leq 0.1$.
		We find that a minimum of $\Delta T_{loop}\sim 0.06$ for $b=2$ is required to lower the scale of the first KK gauge boson below 3 TeV. As $b$ increases corresponding to Higgs moving further towards the IR brane, one can expect minimum $\Delta T_{loop}$ required to keep the KK scale below 3 TeV to increase. The fit values for the input and output observables is given in Table.[\ref{custodialRSout}].

		Fine tuning: In this case too we find that the best fit to the precision data is when the brane mass parameter $b=2$. The KK mass is lowered to 3 TeV thus reducing the fine tuning
		by an order of magnitude. As a result the cancellation between $m_{brane}$ and $m'_{brane}$ is of the order of $0.002$

		\section{Conclusions}
		\label{section6}
		There is a perception that in order that the Randall-Sundrum model
		successfully address the gauge-hierarchy problem the Higgs ought to
		be localised on the IR brane. It has been noted earlier \cite{Luty:2004ye,Quiros:2013yaa} that
		this is not the case and our results bear this out. In fact, we find
		that even if we move the Higgs field off the IR brane, a solution to
		the gauge-hierarchy problem is obtained as long as we have $b\geq2$.
		Further, electroweak fits and fine tuning argument seem to be preferring 
		a $b$ value very close to 2. In the dual CFT terminology, the Higgs field is a partially composite state \cite{batellgherghetta1,batellgherghetta2}. 
		This
		has to do with the exponential form of the scalar profiles which
		get pushed close to the IR brane for values of $b$ greater than 2,
		so that from the point of view of the gauge-hierarchy the Higgs is
		essentially IR-localized. However, such a bulk Higgs differs from
		the brane-localised Higgs in the freedom that it offers in exploring
		the parameter space of the model when confronted with electroweak precision
		constraints. 
		\begin{table}
			\begin{center}
				\begin{tabular}{|c|c|c|}
					\hline
					Model&$m_{KK}(TeV)$&b\\
					\hline
					Normal RS&5.9&2.00\\
					Deformed RS\newline
					($\nu=0.8,\Delta=1$) &2.3&2.00\\
					Deformed RS\newline
					($\nu=1,\Delta=.1$) &1.7&2.00\\
					Custodial RS&2.88&2.00\\
					\hline
				\end{tabular}
			\end{center}
			\caption{Summary of results for the various models at 3$\sigma$}
			\label{conclusions}
		\end{table}
		
		A few remarks about the collider implications of bulk RS models are in
		order. Generically, in these models the gauge boson KK modes provide
		the most interesting signals and the KK gluon is, of these, the most
		important \cite{Agashe:2006hk, Agashe:2007ki, Agashe:2008jb}. The
		production cross-section of the
		KK gauge boson modes is very small partly because of the couplings of 
		these modes to
		the SM particles but also because of the strong constraints on the masses
		of the KK modes coming from electroweak and flavour constraints. The
		cross-sections for other KK modes, like those of the fermions, are
		even smaller than that of the gauge boson KK modes (except in
		some versions of the RS model where the Higgs is treated as a
		pseudo-Nambu Goldstone boson). The collider tests of the bulk
		RS models are therefore difficult and several studies which
		propose probing alternative production channels have been presented
		\cite{Guchait:2007jd, Allanach:2009vz} but the range of masses probed
		by these processes is just
		marginally larger than that allowed by precision constraints. In
		view of this, our results of the global fit for the deformed metric
		case are very encouraging.  Unlike the custodial symmetry case for
		which the global fits yield a bound on the mass of the first KK
		mode of about 2.9 TeV, one gets a lower bound of around 2.3 TeV at 3 $\sigma$
		for the case of the deformed metric for $\nu=0.8~ \text{and}~ \Delta=1$. 
		This bound reduces to about 1.7 TeV $\nu=1 ~ \text{and} ~\Delta=0.1$
		Table[\ref{conclusions}] gives a summary of the results obtained.
		A collider analysis for such class of models was done in \cite{deBlas:2012qf}.
		The deformed metric model
		then is testable at the LHC at a statistically significant level
		and a more detailed study of the collider implications of this
		model is called for.
		
		%In this work we performed a comprehensive analysis to the different versions of the RS model in terms of it's compatibility with precision data. This approach, while being useful
		%to place constraints on the various model parameters, also can serve as an indirect probe for new physics effects in precision experiments in the future. We consider the most
		%general model where all the fields including the Higgs are in the bulk. The localization parameter $b$ for the bulk scalar is closely related to the fine tuning required
		%to generate the light Higgs mass. We find the best fit points to model parameters ($b$ and $\Lambda_{IR}$) which are consistent with electroweak precision tests. Corresponding
		%to these points we evaluate the fine tuning required to generate a light Higgs mass. Three variations of the RS setup were considered:\newline
		%a)RS model with bulk fields\newline
		%b)RS model with a deformed metric\newline
		%c)Model with bulk custodial symmetry. We find that for all three cases, especially the latter two, allowing the input observables to vary helps in reducing the $\Lambda_{IR}$ for the fit.
		%This can have implications for collider searches as

		\textbf{Acknowledgements}
		A.I. and K.S. would like to thank the Centre for High Energy Physics, IISc for its hospitality during his visit where part of the discussions were conducted.

		\bibliographystyle{ieeetr}
		
		\bibliography{RS.bib}
		
	\end{document}